\documentstyle[multicol,aps,epsf]{revtex}
\begin{document}
\title{Stable Distributions in Stochastic Fragmentation}
\author{P.~L.~Krapivsky$^1$, E.~Ben-Naim$^2$, and I.~Grosse$^3$}
\address{$^1$Center for Polymer Studies and Department of Physics,
  Boston University, Boston, MA 02215\\
  $^2$Theoretical Division and Center for Nonlinear Studies,
  Los Alamos National Laboratory, Los Alamos, NM 87545\\
  $^3$Cold Spring Harbor Laboratory, Cold Spring Harbor, NY 11724}
\maketitle
\begin{abstract} 
  
  We investigate a class of stochastic fragmentation processes
  involving stable and unstable fragments. We solve analytically for
  the fragment length density and find that a generic algebraic
  divergence characterizes its small-size tail. Furthermore, the
  entire range of acceptable values of decay exponent consistent with
  the length conservation can be realized. We show that the stochastic
  fragmentation process is non-self-averaging as moments exhibit
  significant sample-to-sample fluctuations.  Additionally, we find
  that the distributions of the moments and of extremal
  characteristics possess an infinite set of progressively weaker
  singularities.
   
\end{abstract}
\begin{multicols}{2}
  
\section{Introduction}
\label{secintro}
         
Fragmentation is a basic stochastic process with a variety of
applications ranging from geology\cite{tur} and fracture\cite{solid}
to the breakup of liquid droplets\cite{liq} and atomic nuclei
\cite{nuc,red}, and impact fragmentation of solid objects
\cite{im,odb,k}.  Fragmentation processes are relevant to numerous
seemingly unrelated problems such as spin glasses, Boolean networks
and genetic populations\cite{sg,h,djm,kauf,flyv,par}. This study is
particularly motivated by computational biology applications involving
partitioning of nonstationary time series into stationary sub-series, and
conversely, reconstruction of a series from its sub-series. Examples
include ``parking'' strategies for genome reconstruction
\cite{rsvth,bat,rts} and DNA segmentation algorithms
\cite{bro,bra,oli}.

The primary feature of the fragmentation process which underlies
segmentation algorithms is that rather than continuing indefinitely,
the process stops when a segment reaches an acceptable ``homogeneity''
level.  Furthermore, fragmentation does not continue indefinitely in
impact fragmentation processes where the kinetic energy and the size
of the fragments determine the number of fragmentation events.  These
applications lead to stochastic fragmentation processes where not only
the way by which fragments are produced, but also the number of
fragmentation events is subject to a random process.

In the simplest stochastic fragmentation process, one starts with a
fragment and breaks it into two pieces.  With probability $p$, a newly
formed fragment remains unstable, i.e, it continues to participate in
fragmentation events, while with probability $q=1-p$, it becomes
stable, and is never fragmented again. The process is repeated for all
unstable fragments until all fragments become stable.  For this simple
process, we show that the final length density, $P(x)$, is purely
algebraic, namely $P(x)=2qx^{-2p}$ \cite{kgb}.  Similar scale free
behavior was observed in other fragmentation
processes\cite{zhang,new}. Stochastic fragmentation also exhibits
intriguing statistical characteristics including moments which are
non-self-averaging, essential singularities in the distribution of
these moments, and an infinite set of singularities in the
distribution of the largest and the smallest fragments.
  
In this paper, we study statistical properties of stochastic
fragmentation processes. In Sec.~\ref{secmodel}, we introduce the
stochastic fragmentation model and derive the fragment length density
$P(x)$. We also consider generalizations to size dependent
fragmentation densities and fragmentation probabilities, and obtain
the exponent underlying the algebraic behavior of $P(x)$ as a root of
a transcendental equation. In Sec.~\ref{secmoments}, we show that the
stochastic fragmentation process is non-self-averaging.  Specifically,
the moments $Y_\alpha=\sum_i x_i^{\alpha}$ exhibit significant
sample-to-sample fluctuations. In Sec.~\ref{secextreme}, we study
extremal characteristics such as the distribution of the largest and
the shortest fragment, and we find that both the extremal
distributions and the distribution of the moments are characterized by
an infinite set of progressively weaker singularities.  In
Sec.~\ref{secrsa}, we present an application to random sequential
adsorption processes, and we summarize this work in
Sec.~\ref{secsummary}.
  
\section{The Model}
\label{secmodel}

In the basic stochastic fragmentation process, we start with the unit
interval, which is considered to be unstable. This interval is
fragmented into two pieces of length $l$ and $1-l$, where $l$ is drawn
from a uniform probability density $\rho(l) = 1$.  With probability
$p$, each of these fragments remains unstable, while with probability
$q \equiv 1-p$, it becomes stable and does not undergo further
fragmentation.  The process is iterated for unstable
fragments until all fragments become stable.

The average total number of stable fragments, $\langle N\rangle$, can
be directly evaluated.  Consider a fragment produced in the first
fragmentation event.  With probability $q$ it is stable, and
consequently, only a single fragment is produced; otherwise, the
process is repeated. Hence $\langle N\rangle= 2(q+p\langle
N\rangle)$, yielding
\begin{equation}
\label{number}
\langle N\rangle=\cases{2q/(1-2p),  & if $p<1/2$;\cr
         \infty,    & if $p\geq 1/2$.\cr}
\end{equation}
The average total number of fragments diverges as the probability $p$
approaches the critical point $p_c=1/2$, reflecting the critical
nature of the corresponding branching process \cite{Harris}.

Next, consider the fragment length density, $P(x)$, of stable
fragments of length $x$.  The recursive nature of the process can be
used to obtain the governing equation for the fragment length density
\begin{equation}
\label{pxeq}
P(x)=2\left[q+p\int_x^1 {dy\over y}\,P\left({x\over y}\right)\right].
\end{equation}
The first term accounts for stable fragments produced at the first
generation, and  the second term describes creation of an $x$-fragment from
a larger $y$-fragment. 

Equation (\ref{pxeq}) can be solved by employing the Mellin transform
technique. Let 
\begin{equation}
\label{mellin}
M(s) \equiv \int dx\, x^{s-1} P(x).
\end{equation}
Hereinafter, integration with unspecified limits is  carried
over the unit interval.  Equations (\ref{pxeq}) and (\ref{mellin})
yield \hbox{$M(s)=2s^{-1}\left[q+pM(s)\right]$}, and consequently, the
Mellin transform of the length density reads
\begin{equation}
\label{ms}
M(s)={2q\over s-2p}.
\end{equation}
Note that the total length is conserved, $M(2)=1$, and that the total
number of stable fragments, $M(1) = \langle N\rangle$, is consistent
with Eq.~(\ref{number}).

The length density can be obtained by inverting the Mellin transform 
\begin{equation}
\label{px}
P(x)=2q\,x^{-2p}.
\end{equation}
Remarkably, the length density is purely algebraic over the entire
range $0<x<1$.  Generally, given an algebraic divergence near the
origin, $P(x)\sim x^{-\gamma}$, length conservation provides the upper
bound $\gamma<2$, and as $0<\gamma=2p<2$, the entire range of possible
divergences is realized by tuning $p$. 

Interestingly, scale free distributions were also found for a dual
stochastic aggregation process where aggregates may turn stable after
each aggregation event. In that case, the large size tail of the
distribution decays algebraically \cite{kbagg}.  We also note that
algebraic distributions have been observed in a number of recent
impact fragmentation experiments involving rods, spheres, bricks,
etc., with the corresponding decay exponents typically ranging between
$1$ and $2$\cite{im,odb,k}.

Notice that the form of the fragment length density is not
altered as the critical point $p_c={1\over 2}$ is passed.
Nevertheless, this point is characterized by a unique property.
Starting from an interval of length $L_0$, Eq.~(\ref{px}) can be
generalized to yield
\begin{equation}
P(x)=2q L_0^{2p-1}x^{-2p}.
\end{equation}
Hence, at the critical point, $P(x)$ becomes independent of the
initial interval length $L_0$.

In the above basic model both the fragmentation density and the
fragmentation probability were independent of the fragment length. In
the following, we show that even when these functions become
length-dependent, $P(x)$ remains algebraic in the small size limit.

\subsection{Arbitrary fragmentation density $\rho(l)$} 

Consider a fragmentation process in which an interval is broken into
two fragments of relative lengths $l$ and $1-l$ with an arbitrary
fragmentation density $\rho(l)$. This density satisfies the
normalization constraint $\int dl\,\rho(l)=1$ and the symmetry
requirement $\rho(l)=\rho(1-l)$. The governing equation for the
fragment length density reads
\begin{equation}
\label{pxeqrho}
P(x)=2q\rho(x)+2p\int_x^1 {dy\over y}\,\rho(y)P\left({x\over y}\right).
\end{equation}
The Mellin transform (\ref{mellin}) of the length density satisfies
$M(s)=2\mu(s)\left[q+pM(s)\right]$, and consequently
\begin{equation}
\label{msrho}
M(s)={2q\over \mu^{-1}(s)-2p},
\end{equation}
where $\mu(s) \equiv \int dl\,l^{s-1} \rho(l)$ denotes the Mellin
transform of the fragmentation density $\rho(l)$.  The symmetry of the
fragmentation density implies $\mu(2)=1/2$ and therefore $M(2)=1$,
which confirms the conservation of length. Also, the normalization
condition implies $\mu(1)=1$, which confirms that the average total
number of fragments is given by $M(1) = \langle N\rangle$, in
agreement with Eq.~(\ref{number}).

The Mellin transform shows that the fragment length density is scale
free only when the fragmentation density is uniform. Nevertheless, the
algebraic small-size behavior remains robust. Indeed, 
Eq.~(\ref{msrho}) suggests that the most important property of $M(s)$
is a simple pole whose location $s=\gamma$ is found from relation
$2p\mu(\gamma)=1$. This simple pole implies a power-law asymptotics of
the fragment length density
\begin{equation}
\label{pxg}
P(x)\simeq Ax^{-\gamma}
\end{equation}
as $x\to 0$.  The exponent $\gamma$ can be determined from the entire
fragmentation density $\rho(x)$ via the relation 
\begin{equation}
\label{gamma}
2p\int dl\, l^{\gamma-1}\rho(l) =1. 
\end{equation}
The prefactor in Eq.~(\ref{pxg}) reads \hbox{$A \equiv
  [q\mu(\gamma)]/[p\mu'(\gamma)]$}, which is simply the residue of the
pole at $s=\gamma$.

 At the critical point $p_c=1/2$ one has $\gamma=1$,
independent of the fragmentation density $\rho(l)$.  The relation
(\ref{gamma}) also shows that a generic behavior $\gamma\to 2$ occurs
if the probability of becoming stable vanishes, i.e., if $p\to 1$. In
the complementary $p\to 0$ limit, the small-size behavior of $\rho(l)$
determines the small fragment distribution. In particular, if
$\rho(l)\sim l^{-r}$ in the limit $l\to 0$, Eq.~(\ref{gamma}) shows
that $\gamma \to r$ as $p\to 0$.  Hence, in this case the restricted 
exponent range $r<\gamma<2$ emerges by tuning $p$.

As an illustration, consider the fragmentation density
$\rho(l)=B[l(1-l)]^{\delta-1}$, with 
$B=\Gamma(2\delta)/\Gamma^2(\delta)$ ensuring proper normalization.
The exponent $\gamma$ is determined from Eq.~(\ref{gamma}) to give 
\begin{equation}
\label{sgamma}
2p\,{\Gamma(2\delta)\,\Gamma(\gamma+\delta-1)\over
\Gamma(\delta)\,\Gamma(\gamma+2\delta-1)}=1. 
\end{equation}
This relation shows that the exponent $\gamma$ always belongs to the
range $1-\delta<\gamma<2$. In the extreme case of $\delta\to 0$, the
decay exponent is concentrated near $\gamma=1$.  Such universal
$P(x)\sim x^{-1}$ behavior is empirically observed in DNA segmentation
algorithms\cite{ivo}.  In the other extreme $\delta\to\infty$, the
exponent simplifies to $\gamma=1+\ln 2p$.  Note also that explicit
results for both $P(x)$ and the exponent $\gamma$ can be obtained for
integer $\delta$'s. The case of $\delta=1$ corresponds to the uniform
density.  For $\delta=2$, the Mellin transform reads
$\mu(s)=6/(s+1)(s+2)$, and Eq.~(\ref{msrho}) yields the
following fragment density
\begin{equation}
\label{pxex2}
P(x)={12q\over \sqrt{1+48p}}\left\{x^{3-\sqrt{1+48p}\over 2}-
x^{3+\sqrt{1+48p}\over 2}\right\}.
\end{equation}
Generally, the length density is a linear combination of $\delta$
power laws for all integer $\delta$'s.

\subsection{Arbitrary fragmentation probability $p(x)$}
 
We now discuss the complementary generalization, in which the
probability $p(x)$ that a new fragment remains unstable depends on the
fragment size $x$. This may actually be the case in impact
fragmentation as well as in DNA segmentation, where fragments have an
intrinsic size scale below which the fragmentation probability becomes
negligible.  For an arbitrary  fragmentation
probability $p(x)$ the governing equation reads
\begin{equation}
\label{pxeq1}
P(x)=2\left[1-p(x)+\int_x^1 {dy\over y}\, p(y) P\left({x\over y}\right)\right]. 
\end{equation}
Consequently, the Mellin transform of the length density admits the
general solution 
\begin{equation}
\label{mspx}
M(s)={2\over s}{1-s\sigma(s)\over 1-2\sigma(s)}
\end{equation}
where $\sigma(s) \equiv \int dx\,x^{s-1}p(x)$ is the Mellin transform
of the probability $p(x)$.

The small size tail is determined by the poles of $M(s)$. When the
condition $2\sigma(0)>1$ is satisfied, $M(s)$ has a simple pole at
$s=\gamma$, and therefore, the small size behavior remains algebraic
as in Eq.~(\ref{pxg}). The corresponding exponent $\gamma$ is
determined from $2\sigma(\gamma)=1$, or explicitly
\begin{equation}
\label{gamma1}
2\int dx\, x^{\gamma-1}p(x)=1, 
\end{equation}
and the prefactor $A=[\gamma-2]/[2\gamma\sigma'(\gamma)]$ equals the
residue at $s=\gamma$. Equation (\ref{gamma1}) is a transcendental
equation, and the details of the function $p(x)$ in the entire range
$0<x<1$ determine the exponent $\gamma$.  This situation is
reminiscent of the behavior found when the fragmentation probability
along the interval was not uniform.  

In the complementary case of $2\sigma(0)<1$, Eq.~(\ref{mspx}) shows
that $M(s)$ has a simple pole at $s=0$, viz.
$M(s)=2s^{-1}/[1-2\sigma(0)]+\ldots$ This implies that the length
density is regular in the small size limit, $P(x)\to 2/[1-2\sigma(0)]$
as $x\to 0$. In the marginal case $2\sigma(0)=1$, we find $M(s)\sim
s^{-2}$, which leads to a logarithmic divergence of the length
density, i.e., $P(x)\sim \ln {1\over x}$ as $x\to 0$.

As an illustration, consider the solvable example $p(x)=x^{\lambda}$
with $\lambda>0$.  In this case $\sigma(s)=1/(s+\lambda)$, and
consequently $M(s)=2\lambda/[s(s+\lambda-2)]$. Inverting the Mellin
transform $M(s)$ gives the fragment length density
\begin{equation}
P(x)=\cases{
{2\lambda\over 2-\lambda}\left(x^{-(2-\lambda)}-1\right)&$0<\lambda<2$;\cr
4\ln {1\over x}&$\lambda=2$;\cr
{2\lambda\over \lambda-2}\left(1-x^{\lambda-2}\right)&$2<\lambda$.\cr}
\end{equation}
Again, the range $0<\gamma<2$ becomes accessible.

Hence, the algebraic small size divergence is robust as it extends to
situations where either the fragmentation density or the fragmentation
probability are size dependent. The entire form of these functions is
needed to calculate the corresponding power-law exponent. 

In the rest of this paper, we restrict ourselves to the basic model
where the fragmentation density is uniform, $\rho(x)=1$, and the
fragmentation probability $p(x)\equiv p$ is size independent.

\section{The Moments} 
\label{secmoments}

The fragment size distribution represents an average over infinitely
many realizations of the stochastic fragmentation processes, and
hence, it does not characterize sample-to-sample fluctuations. In this
section, we show that fluctuations do not vanish in the thermodynamic
limit, and therefore, the process is non-self-averaging.  We
investigate sample-to-sample fluctuations by computing the moments
$Y_\alpha$ defined by
\begin{eqnarray}
\label{defY}
Y_\alpha=\sum_i x_i^\alpha,
\end{eqnarray}
where the sum runs over all fragments in a given realization. These
moments have proved useful in a variety of contexts including spin
glasses, random maps, and random walks \cite{df,der}.  The fact that
these moments have non-trivial probability distributions is a
signature of lack of self-averaging.

Before we attempt to derive these probability distributions, we start
with the simpler task of computing the expected values of the moments
$\langle Y_\alpha \rangle$ as well as their correlations $\langle
Y_\alpha Y_\beta\rangle$.  For integer $\alpha$, $\langle Y_\alpha
\rangle$ is the probability that $\alpha$ randomly
chosen points in the unit interval belong to the same fragment.  Similarly,
for integer $\alpha$ and integer $\beta$, $\langle Y_\alpha
Y_\beta\rangle$ is the probability that among $\alpha+\beta$
points chosen at random, the first $\alpha$ points all lie on the same
fragment, and the last $\beta$ points all lie on another (possibly the
same) fragment.

The expected value of $Y_\alpha$ satisfies
\begin{equation} 
\label{Yav}
\langle Y_\alpha \rangle=\left(q+p\langle Y_\alpha \rangle\right) 
\int dy \left[y^\alpha+(1-y)^\alpha\right].
\end{equation}
The $q$-term corresponds to the situation where a first generation fragment
becomes stable while the second term describes the complementary
situation. Equation (\ref{Yav}) gives
\begin{equation}
\label{Yaver}
\langle Y_\alpha \rangle={2q\over \alpha+1-2p}
\end{equation}
if $\alpha>2p-1$, and $\langle Y_\alpha \rangle=\infty$ if $\alpha\leq
2p-1$. Since the single point averages are simply the moments,
$\langle Y_\alpha \rangle=M(\alpha+1)$, Eq.~(\ref{Yaver}) indeed
agrees with Eq.~(\ref{ms}).

Higher-order averages cannot be computed from the fragment size
density.  However, one can obtain exact expressions for higher averages
from recursion relations similar in spirit to
Eq.~(\ref{Yav}).  For instance, $\langle Y_\alpha Y_\beta\rangle$
satisfies
\begin{eqnarray}
\langle Y_\alpha Y_\beta\rangle&=&2(q+p\langle Y_\alpha Y_\beta\rangle) 
\int dy\,y^{\alpha+\beta}\\
&+&2(q+p\langle Y_\alpha \rangle )(q+p\langle Y_\beta \rangle )\int dy\,
y^\alpha(1-y)^\beta.\nonumber
\end{eqnarray}
The solution to the above recursion relation reads
\begin{eqnarray}
\label{yab}
\langle Y_\alpha Y_\beta\rangle=2\,{q+C(\alpha,\beta)
(q+p\langle Y_\alpha\rangle)(q+p\langle Y_\beta\rangle)\over \alpha+\beta+1-2p}
\end{eqnarray} 
if $\alpha,\beta,\alpha+\beta>2p-1$, and $\langle Y_\alpha
Y_\beta\rangle =\infty$ otherwise.  Here we used the shorthand
notation $C(\alpha,\beta)={\Gamma(\alpha+1)\Gamma(\beta+1)\over
  \Gamma(\alpha+\beta+1)}$. 

Equation (\ref{yab}) shows that $\langle Y_\alpha
Y_\beta\rangle\neq\langle Y_\alpha\rangle\langle Y_\beta \rangle$ and,
in particular, $\langle Y^2_\alpha \rangle\neq\langle
Y_\alpha\rangle^2$. Hence, fluctuations in $Y_\alpha$ do not vanish in
the thermodynamic limit, which implies that the stochastic
fragmentation process is non-self-averaging.  This means that
statistical properties obtained by averaging over all realizations are
insufficient to probe sample-to-sample fluctuations.  Lack of
self-averaging was also found in fragmentation processes that exhibit a
shattering transition \cite{maslov,esipov,olla}.

It is possible to evaluate higher-order averages such as $\langle
Y_\alpha^n\rangle$.  However, even for small $n$ these averages become
quite cumbersome and not terribly illuminating.  Instead, one might
try to obtain the distribution, $Q_\alpha (Y_\alpha)$, of possible
outcomes of the moments $Y_\alpha$.  Let us first consider the
fragment number distribution $Q_0(N)$ (the zeroth moment equals the
number of fragments, $Y_0=N$), which can be determined analytically.
The minimal number of fragments is produced when both of the
first generation fragments are stable, and hence, $Q_0(2)=q^2$.
Similarly for $N\geq 3$ we obtain the recursion relation
\begin{eqnarray}
\label{Q0}
Q_0(N)&=&2pqQ_0(N-1)\\
&+&p^2\sum_{N_1+N_2=N}Q_0(N_1)Q_0(N_2),\nonumber
\end{eqnarray}
where the total number of fragments $N$ is obtained in various ways from
a smaller number of fragments that appear after fragmentation of the two
first generation fragments. Specifically, if exactly one of the first 
generation fragments is unstable, it should produce $N-1$ stable
fragments. If both of the first generation fragments are unstable, they can
produce $N_1$ and $N_2$ fragments, respectively, subject to the
constraint $N_1+N_2=N$.  This explains the right-hand side of
Eq.~(\ref{Q0}).

Equation (\ref{Q0}) can be solved by introducing the generating
function $Q_0(z) \equiv \sum_{N\geq 2} Q_0(N)z^N$, which satisfies
$p^2Q_0^2(z)+(1-2pqz)Q_0(z)+q^2z^2=0$.  Solving this quadratic
equation yields the generating function
\begin{equation}
\label{Q0Z}
Q_0(z)={1-2pqz-\sqrt{1-4pqz}\over 2p^2}.
\end{equation}
Expanding $Q_0(z)$ in powers of $z$ gives 
\begin{equation}
\label{Q0N}
Q_0(N)={\Gamma\left(N-{1\over 2}\right)\over
  \Gamma\left({1\over2}\right)\Gamma(N+1)}\,{(4pq)^N\over 4p^2}.
\end{equation}

At the critical point, $p_c=1/2$, the number distribution
decays algebraically in the large $N$ limit: 
\begin{equation}
\label{Qalg}
Q_0(N)\sim N^{-3/2}.
\end{equation} 
In the vicinity of $p_c=1/2$, the number distribution attains the 
scaling form
\begin{equation}
\label{Qscal}
Q_0(N)\sim N^{-3/2}\exp\left[-4N(\Delta p)^2\right],
\end{equation}
where $\Delta p=p_c-p$. Hence, below the critical point, the tail of the
number distribution is exponential.

The probability that an infinite number of fragments is produced is
given by 
\begin{equation}
\label{Qin}
Q_0(\infty)=1-\sum_{N=2}^\infty Q_0(N)=1-Q_0(z=1).  
\end{equation}
Below the critical point, the number of fragments remains
finite, i.e., $Q_0(\infty)=0$.  In the complementary case of $p>p_c$, 
with finite probability, an infinite number of fragments is produced 
\begin{equation}
\label{Q0inf}
Q_0(\infty)=1-{q^2\over p^2}.
\end{equation}

The case $\alpha=0$ is unique in the sense that the variable $Y_0=N$
is discrete.  Another special case is $\alpha=1$, where length
conservation dictates $Y_1=1$, and therefore the distribution is 
trivial,  $Q_1(Y_1)=\delta(Y_1-1)$.  When $0<\alpha<1$, then
$Y_\alpha>1$, and $Q_\alpha$ has support in the interval $(1,\infty)$,
as in the case of $\alpha=0$.  Equation (\ref{Q0inf}) then suggests
that for $p>p_c$ the distribution $Q_{\alpha}(Y_\alpha)$ should have a
singular component at $Y_\alpha=\infty$.

In the following, we focus on the more interesting case of $\alpha>1$.
Here, the inequality $Y_\alpha<Y_1=1$ implies that the distribution
$Q_{\alpha}(Y_\alpha)$ has support on the interval $(0,1)$. If both of
the first generation fragments happen to be stable, then
$Y_\alpha=x^{\alpha}+(1-x)^{\alpha}$, where $x$ is chosen uniformly in
the unit interval. Both fragments are stable with probability $q^2$,
and the corresponding contribution to $Q_{\alpha}(Y_\alpha)$, which we
denote by $\Pi_\alpha(Y_\alpha)$, reads
$\Pi_\alpha(Y_\alpha)=2q^2{dx\over dY_\alpha}$, where $x$ is the
greater of the two roots of equation
$Y_\alpha=x^{\alpha}+(1-x)^{\alpha}$ (the factor 2 accounts for the
smaller root).  Although it is generally impossible to express the
above formula solely in terms of $Y_\alpha$, in some special cases one
can determine $\Pi_\alpha(Y_\alpha)$ explicitly; for example,
$\Pi_2(Y_2)=q^2(2Y_2-1)^{-1/2}$.

Note that the distribution $\Pi_2(Y_2)$ has a singularity at
$Y_2=1/2$, which obviously implies a singularity of $Q_2(Y_2)$ at the
same point.  To understand the origin of this singularity, notice that
when the process ends with two stable fragments, then
$Y_2=x^2+(1-x)^2\geq 1/2$.  Therefore, the behavior of $Q_2(Y_2)$ for
the case of $Y_2<1/2$ is {\em not} affected by realizations with two
final fragments, and this explains the singularity at $Y_2=1/2$. If
the process ends with three stable fragments, then
$Y_2=x_1^2+x_2^2+(1-x_1-x_2)^2\geq 1/3$.  Similarly, if we end with
$k$ stable fragments, then $Y_2\geq 1/k$.  Hence, we anticipate that
the distribution $Q_2(Y_2)$ has singularities at $Y_2=1/k$ for integer
$k\geq 2$.  Similar singularities underlie distributions of moments in
a number of random processes, including random walks, spin glasses,
random maps, and random trees \cite{h,djm,der,df}.

A straightforward generalization of the above argument suggests that
for arbitrary $\alpha>1$, the distribution $Q_{\alpha}(Y_\alpha)$
possesses singularities at $Y_\alpha=k^{1-\alpha}$. The existence of
these infinitely many singularities shows that analytical
determination of the distribution $Q_{\alpha}(Y_\alpha)$ is hardly
possible.  Indeed, $Q_{\alpha}(Y_\alpha)$ satisfies the difficult
integral equation
\begin{eqnarray}
\label{QY}
&&Q_\alpha(Y_\alpha)=\Pi_\alpha(Y_\alpha)
+2pq\int {dl\over (1-l)^\alpha}\, 
Q_\alpha\left({Y_\alpha-l^{\alpha}\over (1-l)^{\alpha}}\right)\\
&&+p^2\int_0^{Y_\alpha} dZ\,\int {dl\over l^\alpha(1-l)^\alpha}
Q_\alpha\left({Z\over l^\alpha}\right)
Q_\alpha\left({Y_\alpha-Z\over (1-l)^\alpha}\right).\nonumber
\end{eqnarray}
Equation (\ref{QY}) has been derived by repeating the steps used in the
derivation of Eq.~(\ref{Q0}). The first (second) term on the
right-hand side of Eq.~(\ref{QY}) corresponds to the case where two
(one) of the first generation fragments are  stable.  The third
convolution term describes the alternative case when both of the
first generation fragments are unstable. Note that in addition to the
recursive nature of the process, we have employed extensivity, i.e.,
$\langle Y_\alpha\rangle\propto l^\alpha$, in an interval of length
$l$.

In order to study the small-$Y_\alpha$ behavior of the distribution,
we employ the Laplace transform method. From Eq.~(\ref{QY}),
$R_\alpha(\lambda) \equiv \int_0^1 dY_\alpha\,e^{-\lambda
  Y_\alpha}Q_\alpha(Y_\alpha)$ obeys
\begin{equation}
\label{Main}
R_\alpha(\lambda)=p^2\int_0^1 dl\,
R_\alpha\left[\lambda l^\alpha\right]   
R_\alpha\left[\lambda (1-l)^\alpha\right]+\ldots.  
\end{equation}
In Eq.~(\ref{Main}), we do not write explicitly the Laplace transform
of the first two terms of Eq.~(\ref{QY}), because these two terms
become negligible in case of $Y_\alpha\to 0$. The $Y_\alpha\to 0$
asymptotics of $Q_\alpha(Y_\alpha)$ is reflected by the $\lambda\to
\infty$ asymptotics of $R_\alpha(\lambda)$.  We argue that
$R_\alpha(\lambda)\sim \exp(-A\lambda^\omega)$ with $\omega=1/\alpha$.
Assuming that $\alpha\omega>1$, the above stretched exponential form
of $R_\alpha(\lambda)$ shows that the product $R_\alpha\left[\lambda
  l^\alpha\right] R_\alpha\left[\lambda (1-l)^\alpha\right]$ would
reach a maximum that greatly exceeds $R_\alpha(\lambda)$ at $l=1/2$,
in contradiction with Eq.~(\ref{Main}). Alternatively, if
$\alpha\omega<1$, the above product would reach its maximum at $l=0$
and $l=1$. Then, the integral on the right-hand side of
Eq.~(\ref{Main}) would become $2R_\alpha(\lambda)\int dl\,
R_\alpha\left[\lambda l^\alpha\right]\propto\lambda^{-1/\alpha}
R_\alpha(\lambda)$,  in contradiction with Eq.~(\ref{Main}). Therefore, we
conclude that $R_\alpha(\lambda)\sim \exp\left(-A\lambda^{1/\alpha}\right)$  
as $\lambda\to \infty$. This behavior implies that the distribution
$Q_\alpha(Y_\alpha)$ vanishes according to
\begin{equation}
\label{Qasymp}
Q_\alpha(Y_\alpha) \sim \exp\left(-BY_\alpha^{-{1\over \alpha-1}}\right)
\end{equation}
as $Y_\alpha\to 0$.  Therefore, the distribution $Q_\alpha(Y_\alpha)$
has an essential singularity at the origin, which completes a
countable set of algebraic singularities located at
$Y_\alpha=k^{1-\alpha}$ with $k=2,3,\ldots$

\begin{figure}
\centerline{\epsfxsize=7.5cm\epsfbox{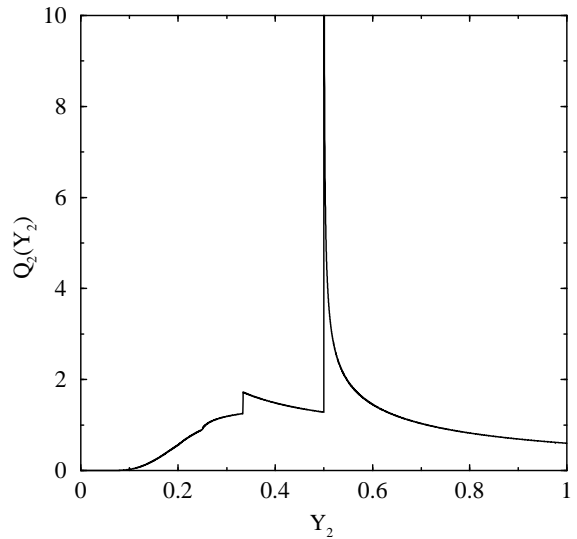}}
\caption{The distribution of the second moment, $Q_2(Y_2)$ versus 
  $Y_2 \equiv \sum_i x_i^2$, from numerical simulations 
  with $p=0.4$.  The data represents
  an average over $5 \times 10^{12}$ realizations.}
\label{figq}
\end{figure}

We performed Monte Carlo simulations of the stochastic fragmentation
process. Hereinafter, we present simulation results for a 
representative case of $p=0.4$. The data corresponds to an average
over $5 \times 10^{12}$ realizations.  Figure \ref{figq} shows the
probability distribution of the second moment.  The distribution
exhibits pronounced singularities at $Y_2=1/2$ and $Y_2=1/3$, while
the following singularities are less visible. One can verify the
existence of further singularities by differentiating $Q_2(Y_2)$.
Figure 2 displays the essential singularity at the origin.

\begin{figure}
\centerline{\epsfxsize=7.5cm \epsfbox{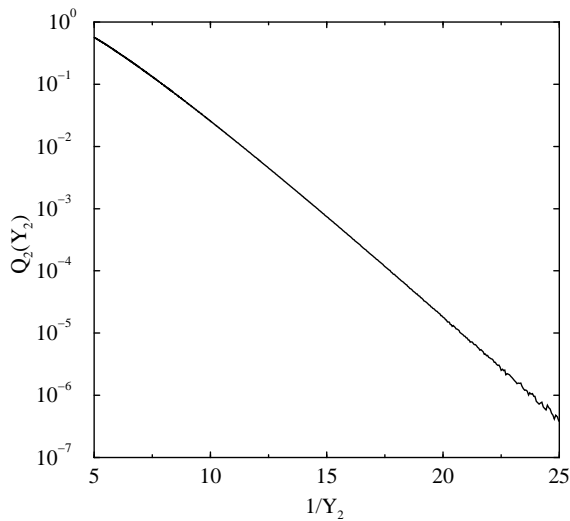}}
\caption{The small size tail of the distribution of the second moment. 
Shown is $Q_2(Y_2)$ versus $1/Y_2$.}
\label{figqtail}
\end{figure}

\section{Extremal Characteristics} 
\label{secextreme}

Extremal properties can be viewed as an additional probe of
sample-to-sample fluctuations.  Moreover, they are interesting on
their own as they arise in many problems of mathematics, physics, and
computer science\cite{gol1,gol2,shepp,knuth,kantor,fik,baik}.  The
largest fragment is an important extremal characteristic.  Obviously,
when $x\geq 1/2$, the size distribution $L(x)$ of the largest fragment
equals the length density, $L(x)=P(x)=2q\,x^{-2p}$.  In the
complementary case of $x<1/2$, $L(x)$ satisfies
\begin{eqnarray}
\label{L} 
L(x)&=&2qpL_-\left({x\over 1-x}\right)+
2p\int\limits_{1-x}^1 {dy\over y}\,L\left({x\over y}\right)\\
&+&2p^2\int\limits_{x}^{1-x} {dy\over y}\,
L\left({x\over y}\right)L_-\left({x\over 1-y}\right),\nonumber 
\end{eqnarray}
where $L_-(u) \equiv \int_0^u dv\,L(v)$. The first term on the
right-hand side of Eq.~(\ref{L}) describes the situation where the
unit interval is fragmented into two intervals of lengths $x$ and
$1-x$, and where the smaller fragment is stable and the larger
fragment is unstable (hence the factor $qp$).  The latter $L_-$ factor
guarantees that subsequent fragmentation of the unstable interval does
not lead to a longer fragment.  If one of the first generation
fragments is shorter than $x$, then only the longest first generation
fragment contributes, which leads to the second term on the right-hand
side of Eq.~(\ref{L}).  The next term describes the situation where
both of the first generation fragments are longer than $x$, so the
longest fragment can result from breaking any of the two fragments.
The factor $L_-$ guarantees that the longest fragment of length $x$
comes from the corresponding first generation fragment, and the factor
$p^2$ guarantees that both of the first generation fragments are
unstable.

\begin{figure}
\centerline{\epsfxsize=7.5cm\epsfbox{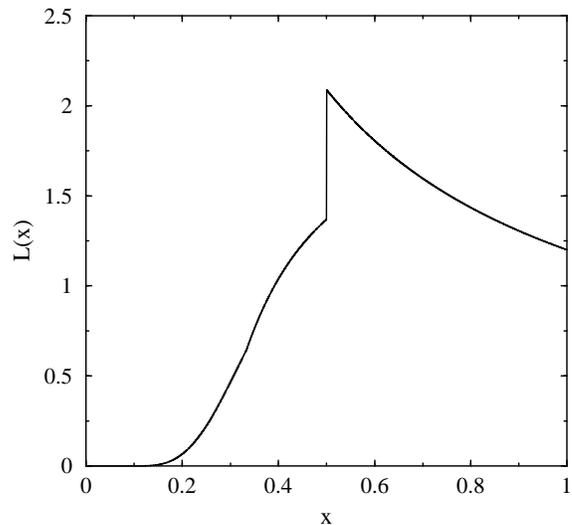}}
\caption{The size distribution of the longest fragment.} 
\label{figl}
\end{figure}

Figure \ref{figl} shows that $L(x)$ is discontinuous at $x=1/2$.  This
can be understood by noting that $L(x)$ obeys different equations for
$x>1/2$ and $x<1/2$, and hence it loses analyticity at the boundary.
Furthermore, we can insert the known result, $L(x)=2q\,x^{-2p}$ for
$x>1/2$, into the right-hand side of Eq.~(\ref{L}) and solve for
$L(x)$ in the interval $1/3<x<1/2$. Then we can use this solution to
determine $L(x)$ in the interval $1/4<x<1/3$, and so on.  Hence,
$L(x)$ should possess an infinite set of singularities at $x=1/k$,
which become weaker as $k$ increases.  One can also understand why
$L(x)$ is discontinuous at $x=1/2$ by considering the $p=0$ case where
the distribution becomes a step function $L(x)=2\theta(x-{1\over 2})$.

Consider now the complementary extremal characteristic --  
the shortest-segment size distribution $S(x)$. Clearly, $S(x)=0$ for
$x>1/2$.  If $x<1/2$, one easily finds $S(x)=2\theta({1\over 2}-x)$ in
the special case of $p=0$.  To proceed in the general case, we first
note that, if the unit interval is divided into $N$ fragments, the
shortest fragment must obey $x_{\rm min}\leq 1/N$.  Hence $1/3<x<1/2$
implies that the unit interval has been divided just once, i.e., both
of the first generation fragments are stable. This shows that
$S(x)=2q^2$ when $1/3<x<1/2$.  Finally, for $x<1/3$ the shortest size
distribution $S(x)$ obeys 
\begin{eqnarray}
\label{S} 
S(x)&=&2q^2+2qpS_+\left({x\over 1-x}\right)\\
&+&2p\int\limits_{x}^{1-x} {dy\over y}\,S\left({x\over y}\right)
\left[q+pS_+\left({x\over 1-y}\right)\right],\nonumber
\end{eqnarray}
where $S_+(u)=\int_u^1 dv\,S(v)$.  The first term on the right-hand
side of Eq.~(\ref{S}) describes the situation where both of the first
generation fragments are stable. The second term corresponds to the
case where the smaller first generation fragment of length $x$ is
stable while the longer fragment is unstable, with the $S_+$ factor
ensuring that subsequent fragmentation of this longer fragment does
not produce a fragment shorter than $x$.  The last term describes
various situations that are possible if both of the first generation
fragments are longer than $x$.

We cannot obtain an analytical expression for $S(x)$ over the entire
length range, because in every interval $({1\over k},{1\over k+1})$ a
different analytical expression holds. In principle, however, one
could determine $S(x)$ recursively.  For instance, we already know
$S(x)$ in the first two regions.  Inserting those expressions into
Eq.~(\ref{S}) yields 
\begin{eqnarray}
\label{S3} 
S(x)=2q^2+4pq^3\left({1\over 2}-{x\over 1-x}+\ln{1-x\over 2x}\right)
\end{eqnarray}
in the third region $1/4<x<1/3$. Clearly, $S(x)$ possesses an infinite
set of singularities at $x=1/k$.

\begin{figure}
\centerline{\epsfxsize=7.5cm\epsfbox{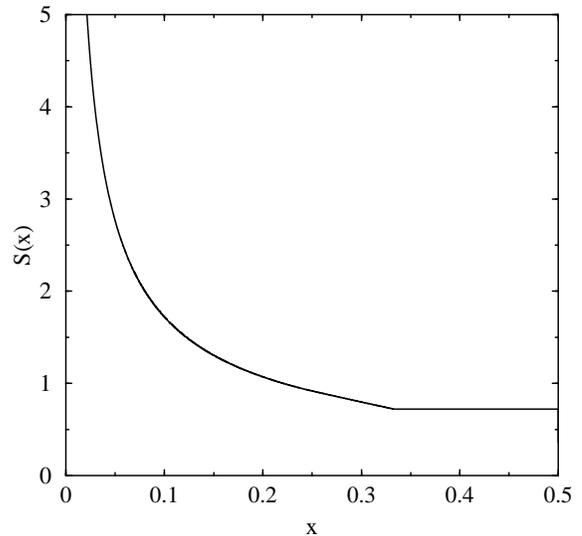}}
\caption{The size distribution of the shortest fragment.} 
\label{figs}
\end{figure}

Figure \ref{figs} shows $S(x)$ for $x<1/2$ and $p=0.4$. One can see the
plateau region $1/3<x<1/2$, and the value of $S(x)$ in this region
agrees with the theoretical prediction $S(x)=2q^2$.  Notice that the
divergence in the small size limit is consistent with the power law
behavior, $S(x)\sim x^{-\delta}$, as $x\to 0$.  Using this
observation, we insert it into Eq.~(\ref{S}), and find a
self-consistent value of the exponent $\delta=2p$. Numerical
simulations show that $S(x)$ slowly approaches the predicted behavior
for the case $p=0.4$ (see Fig.~5).

In deriving the relation $\delta=2p$, we have implicitly assumed that
the shortest-segment size distribution is non-singular. This is indeed
the case when $p\leq p_c$.  For $p>p_c$, however, the distribution
$S(x)$ should additionally contain the singular component
\begin{equation}
\label{Ssing}
S_{\rm sing}(x)=\Delta\delta(x)
\end{equation}
with $\Delta=1-{q^2\over p^2}$, reflecting that with finite
probability, the total number of fragments is infinite, see
Eq.~(\ref{Q0inf}).

A more direct way to derive the same result is to note that $\Delta$,
the probability that $x_{\rm min}=0$, satisfies
$\Delta=2pq\Delta+p^2[1-(1-\Delta)^2]$.  Indeed, the first term
describes the situation when exactly one first generation fragment is
unstable while the second term describes the situation when both of
the first generation fragments are unstable. By solving the above
equation we find two solutions, $\Delta=0$ and $\Delta=(2p-1)/p^2$.
The first solution applies when $p<p_c$; the second solution applies
when $p>p_c$ and agrees with Eq.~(\ref{Ssing}). In order to
investigate the small-size asymptotics of $S(x)$, we write
$S(x)=\Delta\delta(x)+{\cal S}(x)$ and assume that the continuous part
follows the power law behavior, ${\cal S}(x)\sim x^{-\delta}$ as $x\to
0$. Substituting this into Eq.~(\ref{S}) and balancing the dominant 
terms yields $\delta=2q$. To summarize, different behaviors 
characterize the small size tail of $S(x)$
\begin{equation}
\label{sxdif}
S(x)\sim\cases{x^{-2p}&$p<1/2$;\cr
           x^{-2q}&$p>1/2$.}
\end{equation}

\begin{figure} 
\centerline{\epsfxsize=7.5cm\epsfbox{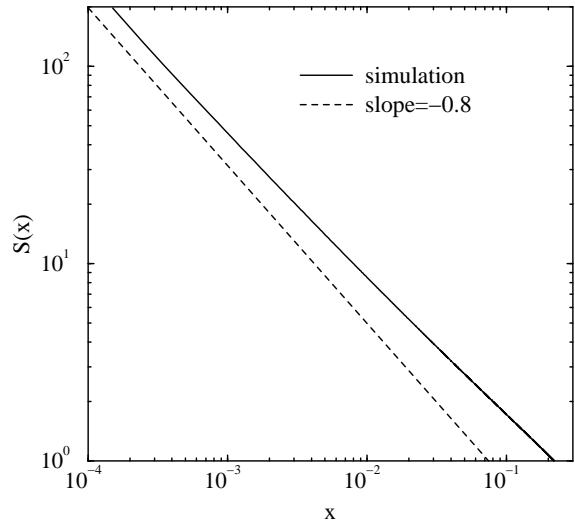}}
\caption{The small size tail of the size distribution of the 
shortest fragment for $p=0.4$. A line of slope $-0.8$ is shown for reference.} 
\label{stail}
\end{figure}

\section{Application to Random Sequential Adsorption}
\label{secrsa}

Random sequential adsorption (RSA) processes\cite{evans} have been
applied to a wide range of chemical, biological, and physical
processes. Examples include binding of proteins to surfaces
\cite{mv,gh}, genome sequencing\cite{rsvth,bat,rts}, and granular
compaction\cite{kbpark,bknjn}.  In one dimension, $k$-mers are deposited
(with a uniform rate) onto a linear lattice, and the deposition events
are successful only when all $k$ sites are empty.  Eventually, the
system reaches a jamming state where no further deposition events are
possible.  Basic quantities of interest are the jamming density,
$\rho_{\infty}$, and its dependence on the initial concentration
$\rho_0$, as well as the gap size distribution.

Adsorption can be viewed as a fragmentation process, and the above
stochastic fragmentation process generalizes RSA to situations where
the gaps between the adsorbed particles may become passive with
probability $p$ after each deposition event.  Stable gaps can no
longer be deposited onto, or in other words, fragmented. In the
following, we study this RSA problem both in discrete and continuous
space. We show that the limiting density of passive gaps is
proportional to the fragment length density obtained above.  In the
interesting limit of vanishing initial concentrations, $\rho_0\to 0$,
the final jamming density vanishes according to $\rho_{\infty}\sim
\rho_0^{2q}$  $q<1/2$, and as $\rho_{\infty}\sim \rho_0$ if $q>1/2$.
Hence, the jamming density is significantly enhanced over the initial
density only if $p$ exceeds the critical value $p_c=1/2$.

\subsection{Discrete Space}

Let us consider random sequential adsorption in one spatial dimension
were each deposition event creates two new smaller gaps. We assume
that each of these gaps remains active with probability $p$, while
with probability $q=1-p$ it becomes passive, i.e., adsorption events
no longer occur on this gap. Initially, the system consists of
randomly distributed monomers with density $\rho_0$, and all gaps are
active.  Then, $r$-mers are deposited with a temporally constant and
spatially homogeneous rate, set to unity without loss of generality.
The deposition events are successful if and only if all $r$ sites are
empty.  Below, we consider the monomer case ($r=1$).

Let $A_k(t)$ ($P_k(t)$) denotes the probability of finding an active
(passive) gap of size $k$ at time $t$. These gap probability
distributions satisfy the following master equations
\begin{eqnarray}
\label{ap} 
{dA_k(t)\over dt}&=&2p\sum_{m=k+1}^\infty A_m(t)-kA_k(t),\nonumber\\
{dP_k(t)\over dt}&=&2q\sum_{m=k+1}^\infty A_m(t).
\end{eqnarray}
The loss term reflects that deposition in each of the empty sites
destroys the gap, and the gain term reflects the fact that two smaller
gaps are created in each deposition event. The prefactors of the gain
term equal the corresponding production rates. It is easy to verify
that the evolution equations conserve the total length $\sum_k (k+1)
(A_k+P_k)=1$.

Let us consider systems that initially consist of randomly distributed
monomers with all gaps being active, i.e.,
$A_k(0)=\rho_0^2(1-\rho_0)^k$ and $P_k(0)=0$. Since $P_k$ is enslaved
to $A_k$, we derive the latter quantity first. The linear loss rate
suggests the exponential ansatz
\begin{equation}
\label{exp}
A_k=\alpha \beta^k
\end{equation}
with the initial values $\alpha(0)=\rho_0^2$ and $\beta(0)=1-\rho_0$.
Substituting (\ref{exp}) into the rate equation yields 
\begin{equation}
\label{ab} 
{d\beta\over dt}  =  -\beta,  \qquad
{d\over dt}\ln \alpha  =  2p\,{\beta\over 1-\beta}.
\end{equation}
The first equation yields $\beta(t)=(1-\rho_0)e^{-t}$, and the second
equation can be conveniently solved by changing the time variable from
$t$ to $\beta$ using $\beta dt=-d\beta$.  This transforms the second
equation into ${d\over d\beta}\ln \alpha=-2p(1-\beta)^{-1}$.
Integrating this equation subject to the above initial condition
yields $\alpha=\rho_0^{2q}(1-\beta)^{2p}$. The time-dependent gap
distribution is therefore
\begin{equation}
\label{ak}
A_k  =  \rho_0^{2q}(1-\beta)^{2p}\beta^k,\qquad 
\beta  =  (1-\rho_0)e^{-t}.
\end{equation}

Next, we study the final jamming density which can be obtained from
the active gap distribution by integration of the overall deposition
rate over time, i.e., $\rho_{\infty}-\rho_0=\int_0^{\infty} dt \sum_k k
A_k(t)= \int_0^{\infty} dt\, \alpha \beta(1-\beta)^{-2}$. Again, it is
useful to transform $t$ to $\beta$. Evaluating the integral gives
\begin{equation}
\rho_{\infty}=\cases{(\rho_0^{2q}-2q\rho_0)/(1-2q)  &$q\ne 1/2$;\cr
\rho_0\ln(1/\rho_0) &$q=1/2$.}
\end{equation}
It is easy to verify that $\rho_{\infty}\to 1$ in the limit $q\to 0$.
In the more interesting limit when the system is initially almost
empty, i.e., when $\rho_0\to 0$, the following leading
behaviors emerge
\begin{equation}
\label{rholim}
\rho_{\infty}\sim\cases{
\rho_0^{2q}           &$q<1/2$;\cr
\rho_0\ln(1/\rho_0) &$q=1/2$;\cr
\rho_0              &$q>1/2$.}
\end{equation}
In other words, the final coverage depends algebraically on the
initial coverage in the range $q<1/2$. In this case, the adsorption
process can be viewed as ``effective'' since the increase in density
is significant. Otherwise, the final density is proportional to the
initial density $\rho_0$. The critical point is marked by a weak
logarithmic increase in the jamming density.

We turn now to the distribution of passive gaps which can be obtained
by integrating the master equation (\ref{ap}) with $A_k(t)$ given by
Eq.~(\ref{ak}). In order to compare the gap-size distribution with the
fragmentation case, we focus on the limiting ($t\to\infty$)
distribution of passive gaps, which reads
\begin{equation}
\label{Pinf}
P_k(\infty)=2q\rho_0^{2q}\int_0^{1-\rho_0} d\beta\, (1-\beta)^{2p-1}\beta^k.
\end{equation}
In the limit of almost empty initial conditions, $\rho_0\to 0$, where
the average gap size diverges, we find $\langle k\rangle
=(1-\rho_{\infty})/\rho_{\infty}\sim \rho^{-1}_{\infty}$.

Hence, the most interesting behavior emerges in the scaling region
$\rho_0\to 0$ and $k\to\infty$ with the scaling variable $\xi=k\rho_0$
kept fixed.  In this region the limiting gap distribution
(\ref{Pinf}) can be rewritten in the scaling form
\begin{equation}
\label{scal} 
P_k(\infty)=2q\rho_0^{2q}k^{-2p}\Gamma(2p,\xi),
\end{equation}
where $\Gamma(a,\xi)=\int_{\xi}^{\infty} dx\, x^{a-1}e^{-x}$ denotes the
incomplete gamma function. Equation (\ref{scal}) shows that the gap size
distribution behaves algebraically as long as the size of the gap does
not exceed the average initial size $k^*=\rho_0^{-1}$, while for larger
gaps the size distribution is suppressed exponentially 
\begin{equation}
\label{aklim}
P_k(\infty)\simeq\cases{
2q\Gamma(2p)\rho_0^{2q}k^{-2p}  &$k\ll k^*$;\cr
2q\rho_0k^{-1}e^{-k/k^*}&$k\gg k^*$.}
\end{equation}
These two expressions are indeed of the same order,
$P\sim\rho_0^2$, in the vicinity of the crossover point $k\sim k^*$.
In general, we find the algebraic behavior $P_k(\infty)\sim k^{-2p}$
in the limit $\rho_0\to 0$, in agreement with the stable
fragment length density of Eq.~(\ref{px}). This agrees with
intuition since in the limit $\rho_0\to 0$ the stochastic RSA is a
discrete counterpart of the stochastic fragmentation.

The above treatment can be generalized to the dimer ($r=2$) case and
even to the general $r$-mer case. Although these solutions become very
cumbersome as $r$ grows, the asymptotic behavior found for the monomer
case including the scaling form of $P_k(\infty)$ and the jamming density
$\rho_{\infty}$ are not altered.

\subsection{Continuous Space}

The continuum limit where particles of unit length are deposited
irreversibly onto a line can be obtained from the discrete $r$-mer
case by taking the limit $r\to\infty$ and by redefining the time
variable $rt \to t$ and the initial density $r \rho_0 \to \lambda$.
Then, the densities of active and passive gaps of size $x$ evolve
according to
\begin{eqnarray}
\label{ratex} 
{\partial A(x,t)\over \partial t}
&=&2p\int_{x+1}^\infty dy\, A(y,t)-\theta(x-1)(x-1)A(x,t),\nonumber\\
{\partial P(x,t)\over \partial t}&=&2q\int_{x+1}^\infty dy\, A(y,t),
\end{eqnarray}
where $\theta(x)$ is the step function.  The initial conditions read
$A(x,0)=\lambda^2 e^{-\lambda x}$ and $P(x,0)=0$.

We first derive the distribution of active gaps of lengths $x\geq
1$. Equation (\ref{ratex}) suggests that it remains exponential
throughout the evolution, i.e., 
\begin{equation}
\label{ax}
A(x,t)=\Phi(t)\exp\left[-\lambda x-(x-1)t\right].
\end{equation}
Substituting this exponential form into Eq.~(\ref{ratex}) yields
${d\over dt}\ln \Phi(t)=2p\,e^{-\lambda-t}/(\lambda+t)$.  Integrating
this differential equation subject to the initial conditions
$\Phi(0)=\lambda^2$ gives the time dependent prefactor 
\begin{equation}
\label{pt}
\Phi(t)=\lambda^{2q}(\lambda+t)^{2p}
\exp\left[-2p\int_\lambda^{\lambda+t}d\tau\,{1-e^{-\tau}\over \tau}\right].
\end{equation}

The jamming density can be obtained by integrating the total deposition
rate over time, $\rho_{\infty}=\int_0^{\infty} dt
\int_1^\infty dx\,(x-1) P(x,t)$, which yields 
\begin{equation}
\rho_{\infty}=\lambda^{2q}e^{-\lambda}\int_\lambda^{\infty} 
{dt\over t^{2q}} 
\exp\left[{-2p\int_{\lambda}^{t} d\tau\,{1-e^{-\tau}\over \tau}}\right].
\end{equation}
When $p=1$, this expression agrees with the jamming density of the
parking model with and without disorder \cite{ren,bk}. Independent of
the probability $p$, the approach to the jamming state follows the
classical $t^{-1}$ law \cite{evans}
\begin{equation}
\rho_{\infty}-\rho(t)\simeq 
\lambda^{2q}e^{-\lambda}(\lambda+t)^{-1}\sim t^{-1}.
\end{equation}
Additionally, the leading behavior in the limit of dilute initial
conditions ($\lambda\to 0$) can be evaluated and the behavior
found in Eq.~(\ref{rholim}) generalizes to the continuum limit 
\begin{equation}
\rho_{\infty}\sim\cases{
\lambda^{2q}           &$q<1/2$;\cr
\lambda\ln(1/\lambda) &$q=1/2$;\cr
\lambda              &$q>1/2$.}
\end{equation}

The limiting passive gap distribution is found by integrating the rate
equation (\ref{ratex}) using the active gap distribution 
\begin{equation}
\label{Pint} 
P_\infty(x)=2q\int_0^{\infty} {dt\over \lambda+t}\,
\Phi(t)e^{-\lambda-(\lambda+t)x},
\end{equation}
with $\Phi(t)$ given by Eq.~(\ref{pt}). In the limit of dilute initial
conditions ($\lambda\to 0$) one can simplify the integral on the
right-hand side of Eq.~(\ref{Pint}) to find the following extremal
behaviors of the gap distribution
\begin{equation}
\label{lim}
P_\infty(x)\simeq\cases{
2q\lambda x^{-1}e^{-\lambda x}   &$x\gg \lambda^{-1}$;\cr
2q\Gamma(2p)\lambda^{2q}x^{-2p}  &$1\ll x\ll \lambda^{-1}$;\cr
2q\,e^{-2p\gamma_E}\lambda^{2q}\,\ln\left({1\over \lambda x}\right)&$x\ll 1$;}
\end{equation}
where $\gamma_E=0.577215\ldots$ denotes the Euler constant. The first
two asymptotics in Eq.~(\ref{lim}) are straightforward extensions of
the corresponding behaviors in the lattice case, and the last line of
Eq.~(\ref{lim}) has been derived from Eq.~(\ref{Pint}) using $\int_0^T
d\tau\,({1-e^{-\tau})/\tau} =\ln T+\gamma_E+{\cal
  O}\left(e^{-T}\right)$.

Finally, we note that even in the long-time limit the density of active
gaps does {\em not} vanish for sufficiently short gaps, $x\leq 1$.
One can determine $A_\infty(x)$, and more generally $A(x,t)$, by
employing an elementary relation between the densities of active and
passive gaps, namely
\begin{equation}
\label{AP}
A(x,t)=A(x,0)+{p\over q}\,P(x,t).
\end{equation}
This relation immediately follows from the master equations
(\ref{ratex}), and it clearly holds for arbitrary $t$ as long as
$x\leq 1$.  By combining (\ref{lim}) and (\ref{AP}) we find
\begin{equation}
\label{limA}
A_\infty(x)\simeq 2p\,e^{-2p\gamma_E}\,\lambda^{2q}\,
\ln\left({1\over \lambda x}\right),
\end{equation}
which applies if $\lambda\ll 1$ and $x\ll 1$. 

Therefore, stochastic fragmentation processes can be naturally
extended to adsorption processes, and apart from numeric prefactors
the algebraic fragment-size distribution is reproduced in the limit of
empty initial conditions. The phase transition underlying the
branching process has an interesting implication. The jamming density
is significantly larger than the initial density only when $p>1/2$.
The super-critical nature of the underlying branching process allows
for an infinite number of fragments produced from a single fragment,
and this explains the enhanced jamming density in the stochastic RSA
process.

Although the fragmentation and the adsorption results are closely
related, we have used two complementary approaches to obtain them. In
the former case, it was convenient to bypass the distribution of
unstable fragments and solve directly for the final stable fragment
distribution, while in the latter case, it was more natural to study
the entire time dependent behavior of both distributions. A more
complete treatment of the fragmentation process is of course possible
using a continuous time formulation which leads to rate equations
similar to Eq.~(\ref{ratex}).

\section{Summary}
\label{secsummary}

We have studied a class of stochastic fragmentation processes, where
fragments may become stable (``frozen'') after each fragmentation
event. We have found that in general, these processes are
characterized by an algebraic small-size divergence of the fragment
size distribution.  This behavior is robust as it holds for size
dependent fragmentation densities and fragmentation probabilities, as
well as in dual adsorption processes in both continuous and discrete
space. The corresponding power-law exponents can be tuned by varying
the fragmentation probability, and the entire range allowed by
mass conservation may be realized.

While the size density can be determined analytically, additional
statistical measures of fluctuations are more difficult to handle.
Nevertheless, we have shown that moments of the distribution exhibit
large sample-to-sample fluctuations, and hence, knowledge of the
entire distribution of observables is needed to characterize the
system.  Additionally, the distribution of the moments and of extremal
characteristics, such as the longest and the shortest fragments,
possesses an infinite set of singularities. 

Lack of self-averaging is important in practical applications such as
utilization of DNA segmentation for comparison of genomes of different
species.  Great care is clearly needed in comparative analysis of the
segment length distributions as observed deviations between segments
may be actually statistical rather than biological.

\bigskip\noindent We thank Michael Zhang for valuable discussions.
This research was supported by DOE (W-7405-ENG-36), NSF (DMR9978902), ARO
(DAAD19-99-1-0173), NIH, and the CSHL Association.

\end{multicols}
\end{document}